# Evaluation of extra pixel interpolation with mask processing for medical image segmentation with deep learning

Olivier Rukundo

Center for Clinical Research, University Clinic of Dentistry, Medical University of Vienna, Vienna, Austria

## ABSTRACT

Current mask processing operations rely on interpolation algorithms that do not produce extra pixels, such as nearest neighbor (NN) interpolation, as opposed to algorithms that do produce extra pixels, like bicubic (BIC) or bilinear (BIL) interpolation. In our previous study, the author proposed an alternative approach to NN-based mask processing and evaluated its effects on deep learning training outcomes. In this study, the author evaluated the effects of both BIC-based image and mask processing and BIC-and-NN-based image and mask processing versus NN-based image and mask processing. The evaluation revealed that the BIC-BIC model/network was an 8.9578 % (with image size 256 × 256) and a 1.0496 % (with image size 384 × 384) increase of the NN-NN network compared to the NN-BIC network which was an 8.3127 % (with image size 256 × 256) and a 0.2887 % (with image size 384 × 384) increase of the NN-NN network.

## 1. INTRODUCTION

The NN interpolation algorithm is often used for interpolation mask processing and other operations, such as in data augmentation, in which geometric transformations are performed. This is due to its inherent advantage of not creating non-original or extra class labels in the interpolated masks [6]. In our previous study, in [6], the author demonstrated that NN-based mask processing could lead to suboptimal deep learning training outcomes. Technically, that could happen because of heavy jagged edges and/or texture resulting from rounding functions-based original pixel replications [7], [9], [10], [11]. Or else, it could simply happen because the NN interpolation does not provide a smooth and accurate representation of object shapes [15]. And, that could lead to overestimating or underestimating the outline or size of objects thus affecting negatively the accuracy of segmentation with deep learning. In this regard, the NN interpolation algorithm could exacerbate the errors resulting from the failure of human or non-human annotators to accurately trace the shape of real edges or boundaries of objects of interest [12], [13], [14]. In the previous study, the author hypothesized the use of extra pixel interpolation with mask processing to improve the accuracy of image segmentation [6]. To verify that hypothesis, the author introduced a novel interpolation mask processing to remove extra class labels in masks processed or interpolated using the extra pixel interpolation algorithms, such as BIC or BIL, which typically utilize a weighted average of neighboring pixels to interpolate or produce non-original pixels, thus smooth and visually good results [8]. Experimentally, the author only evaluated the effects of extra pixel interpolation-based image and mask processing on deep learning training outcomes [6]. In this study, the author leveraged the same hypothesis to conduct and present a thorough evaluation of BIC-based image and mask processing and BIC-and-NN-based image and mask processing versus NN-based image and mask processing. In other words, the thorough evaluation conducted focused on the performance of extra pixel category-based networks/models and the combination of extra and non-extra pixel networks against the non-extra pixel category-based networks. The rest of the paper is organized as follows: Part 2 covers the materials and methods; Part 3 presents the experimental results and relevant discussions; and Part 4 concludes the evaluation study.

## 2. MATERIALS AND METHODS

### 2.1. U-net architecture

U-Net is a CCN architecture widely used for semantic segmentation tasks in bio/medical image analysis [1], [2], [5]. It features a U-shaped design, comprising contracting and expansive paths [6]. The contracting path consists of repeating blocks of convolution, ReLU activation, and max pooling. The expansive path involves transposed convolution, ReLU activation, concatenation with the down-sampled feature map, and additional convolution. Figure 1 shows the input and output layers, as well as the intermediate layers and connections, of a deep learning network as visualized by the *analyzeNetwork* function in MATLAB.

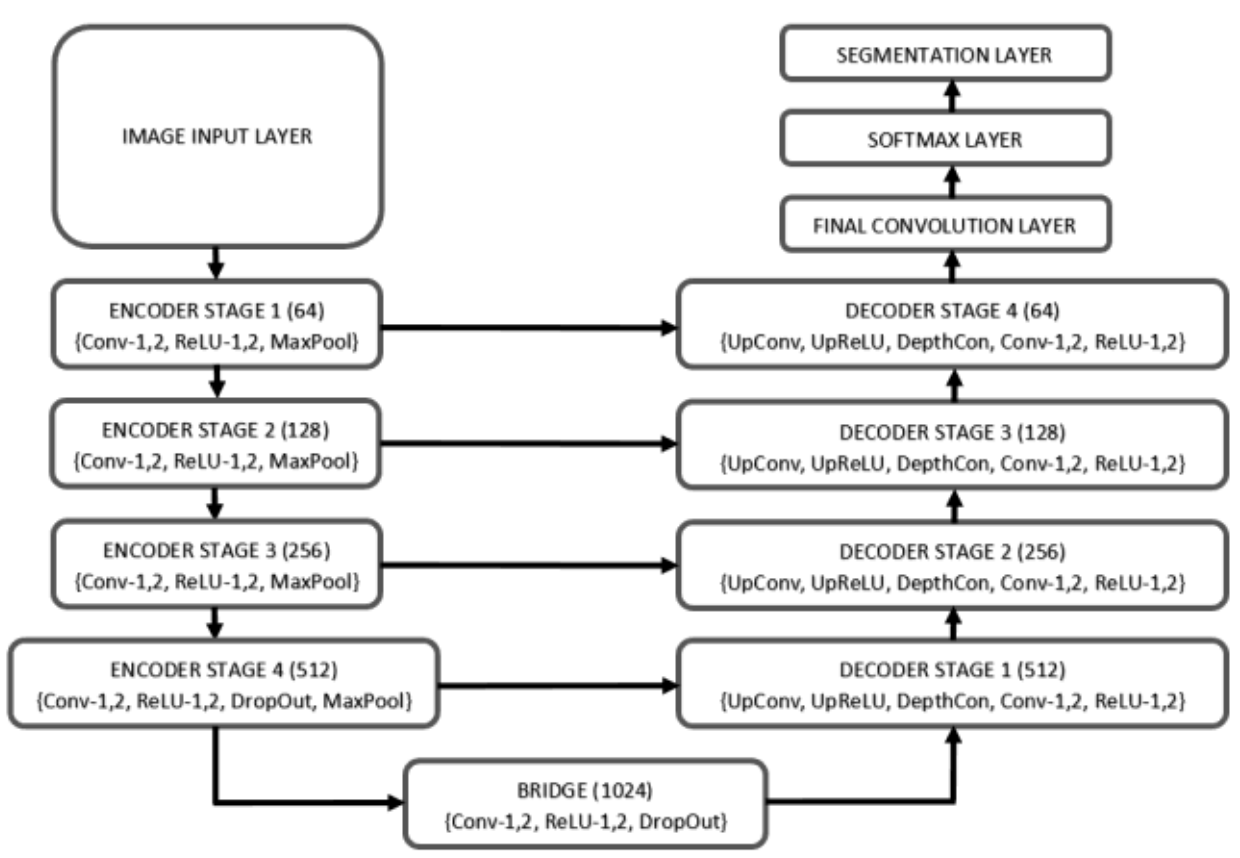

**Figure 1:** U-net architecture. *Conv* means convolution. *ReLU* is a rectified linear unit. *DepthConv* is depth concatenation. *UpConv* means up-convolution or transposed convolution. *MaxPool* is Max Pooling, [6]

Note that, the U-Net Layers function was used to implement the U-net architecture. Also, note that the U-Net Layers function provides options to customize it. For example, in terms of the number of encoder filters (i.e., the number of output channels for the first encoder stage), filter size (the convolution layer kernel or filter size), and encoder depth (i.e., the number of times the input image is downsampled or upsampled during processing). However, due to computing limitations, such customizable options were kept to their default values. For more information, refer to the MATLAB documentation [3] and [4].

### 2.2. Interpolation mask processing

Interpolated mask processing or handling strategy was first proposed in [6] to remove extra class labels in interpolated GT images to keep the number of classes of output mask as input mask. The developed strategy uses three important techniques/operations, namely (1) thresholding, (2) median-filtering, and (3) subtraction to remove extra class in the five steps, as shown in [6]. Figure 2 shows at which stage the interpolation mask processing strategy can be applied to produce the output mask with the same number of classes as the input mask.

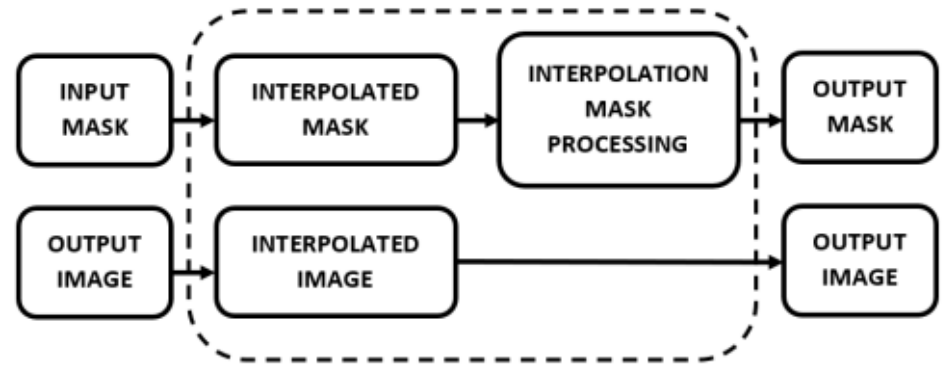

**Figure 2:** Illustration of the application of the interpolation mask processing strategy.

## 3. RESULTS AND DISCUSSIONS

### 3.1. Training data

The reference dataset consisted of 1500 LGE MRI images and corresponding masks of the size 128 × 128. This dataset is a part of the dataset used in [6]. Images and masks were upscaled to create new datasets. Each mask consisted of three classes, with class IDs, corresponding to 255-, 128-, and 0-pixel labels. And, no unacceptable class imbalance was detected in the new datasets. As done in [1] and [2], each training dataset was split into three sets, namely: the training set (60% of the training dataset), the validation set (20% of the training dataset), and the test set (20% of the training dataset). Our U-Net model was trained in a single-GPU environment with the Nvidia GeForce RTX 3070 graphic card and 11th Gen Intel(R) Core(TM) i7-11700F @ 2.50GHz, 2496 MHz, 8 Core(s), 16 Logical Processor(s).

### 3.2. Hyperparameter settings

Here, the author referred to the previous study, in [1], to manually adjust training hyperparameter values, with no new adjustments if 90% of training accuracy was reached during the first 10% of all epochs. Training hyperparameters that were not listed below remained set to default, including the number of first encoder filters, filter size, and encoder depth. The number of epochs = 60; the minimum batch size = 8; the initial learning rate = 0.0001; L2 regularization = 0.000005; optimizer = Adam (adaptive moment estimation algorithm).

### 3.3. Data augmentation

To augment images and their corresponding pixel label images or masks, random reflection and translation transformations were applied. In other words, the data augmentation included the random reflection in the left-right direction and the range of vertical and horizontal translations, with 50% probability, on the pixel interval ranging from -10 to 10.

### 3.4. Loss function

The loss function used was the default cross-entropy function provided by the U-Net Layers function. Note that,

in our U-Net version, the pixel classification layer was not replaced with a weighted pixel classification layer. Further information on this function can be found in the MATLAB documentation [3].

### 3.5. Results

All the images used in our experiments can be seen via the author's GitHub link (https://github.com/orukundo/Extra-Pixel-Interpolation-Experimental-Images). Therefore, they are not included in this section.

#### *3.5.1. Regional networks performance*

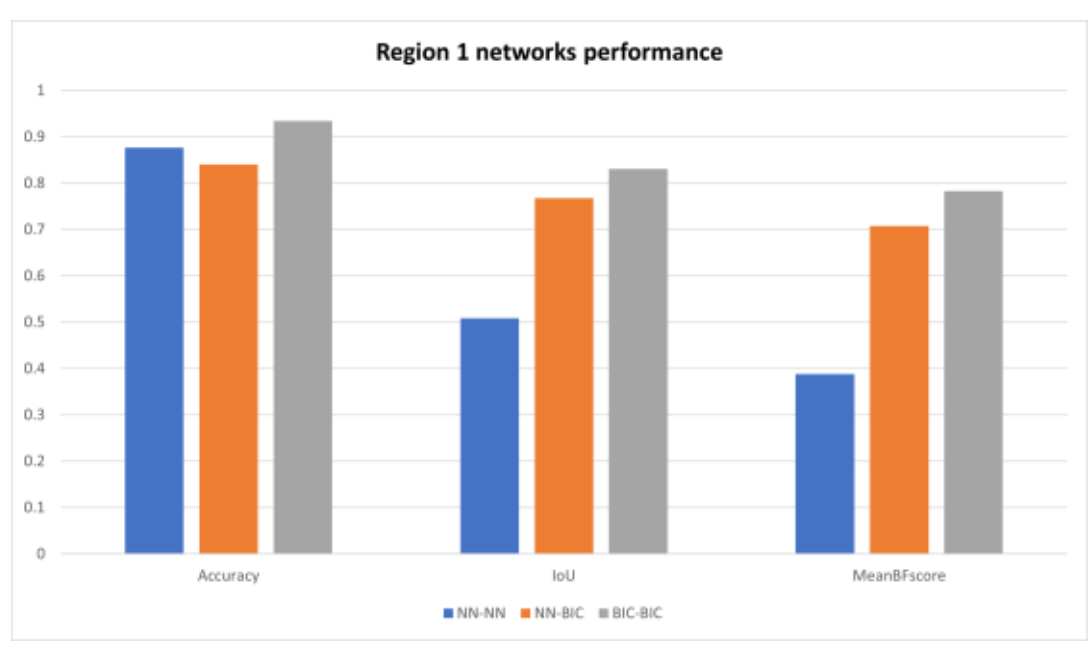

**Figure 3:** Networks trained with datasets consisting of 256 × 256 images and masks: NN-NN, BIC-NN, and BIC-BIC

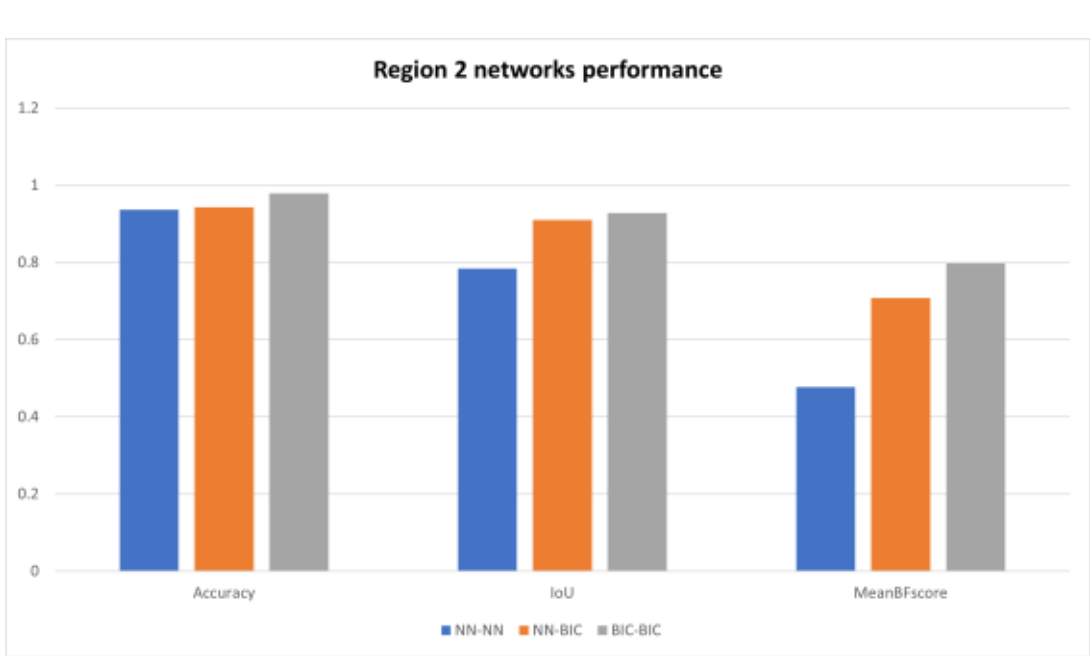

**Figure 4:** Networks trained with datasets consisting of 256 × 256 images and masks: NN-NN, BIC-NN, and BIC-BIC

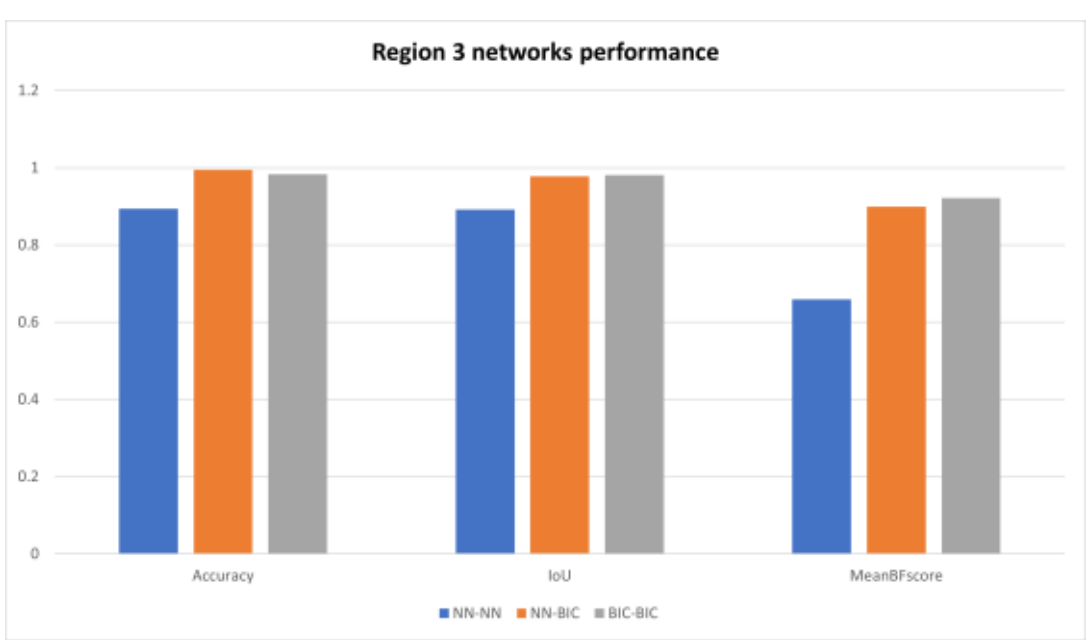

**Figure 5:** Networks trained with datasets consisting of 256 × 256 images and masks: NN-NN, BIC-NN, and BIC-BIC

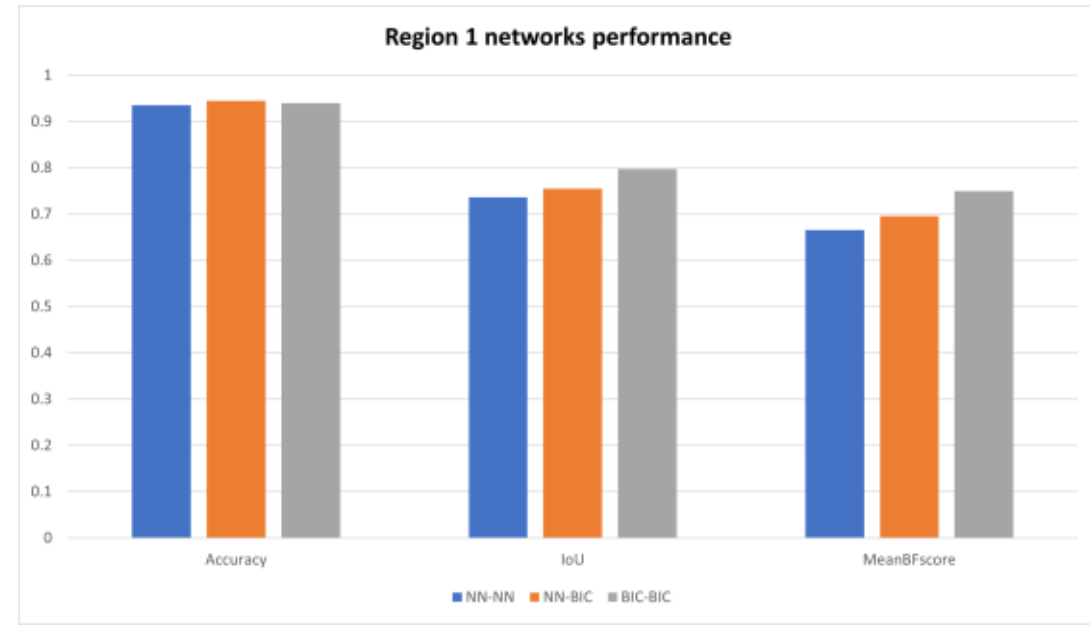

**Figure 6:** Networks trained with datasets consisting of 384 × 384 images and masks: NN-NN, BIC-NN, and BIC-BIC

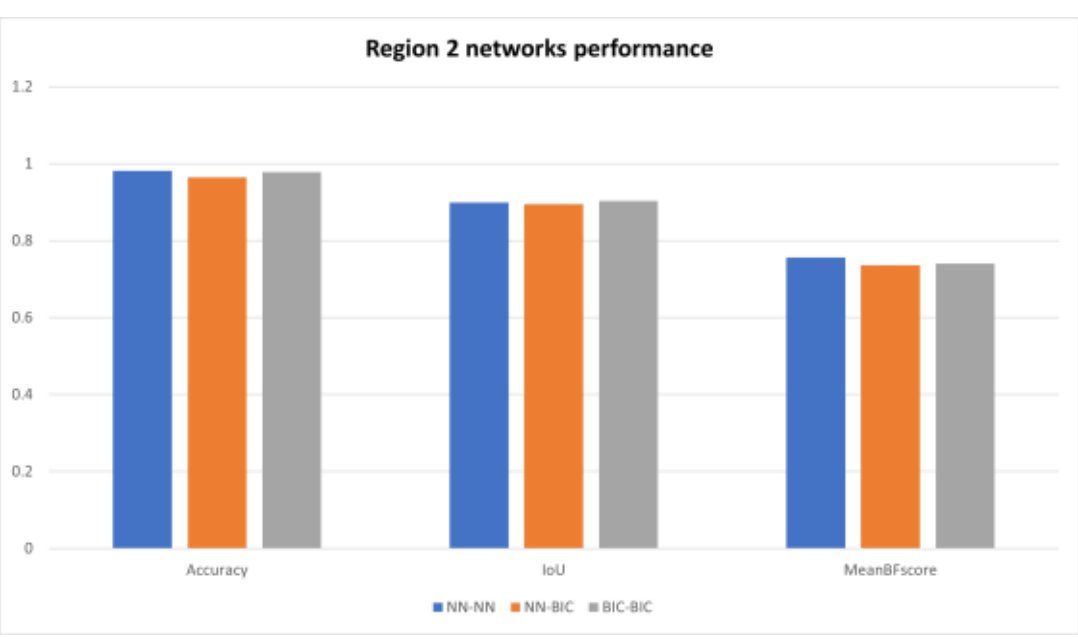

**Figure 7:** Networks trained with datasets consisting of 384 × 384 images and masks: NN-NN, BIC-NN, and BIC-BIC

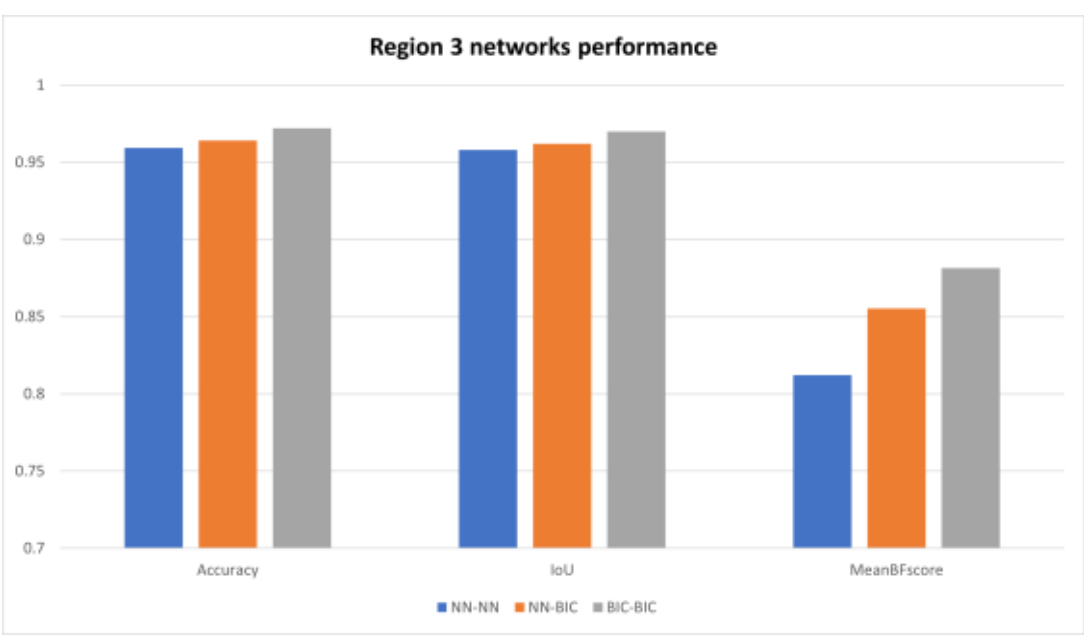

**Figure 8:** Networks trained with datasets consisting of 384 × 384 images and masks: NN-NN, BIC-NN, and BIC-BIC

#### *3.5.2. Global networks performance*

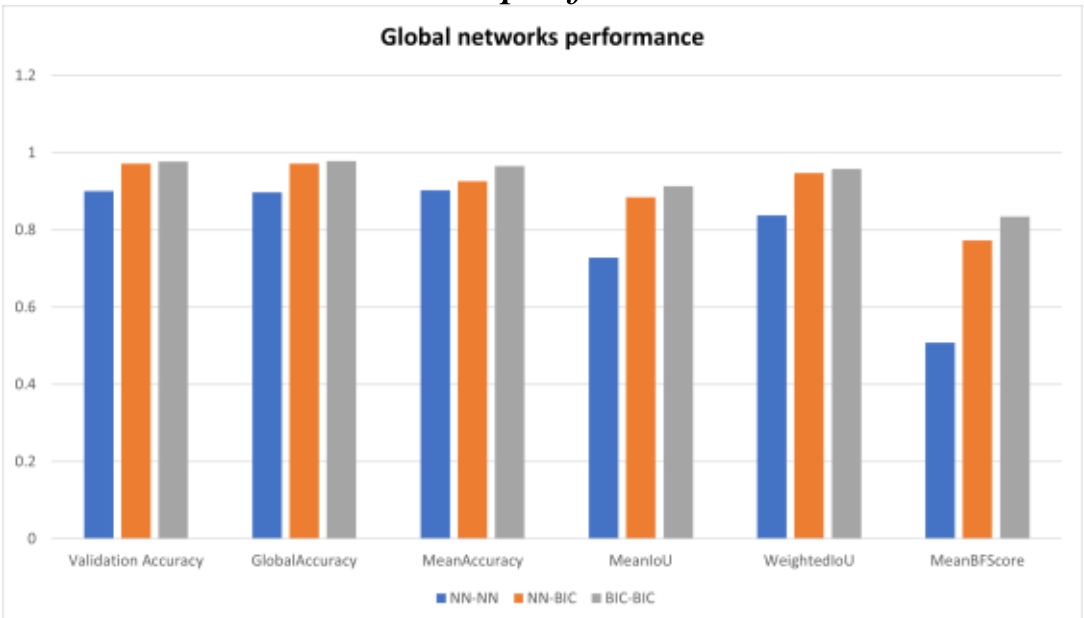

**Figure 9:** Global performance of networks trained with datasets consisting of 256 × 256 images and masks: NN-NN, BIC-NN, and BIC-BIC

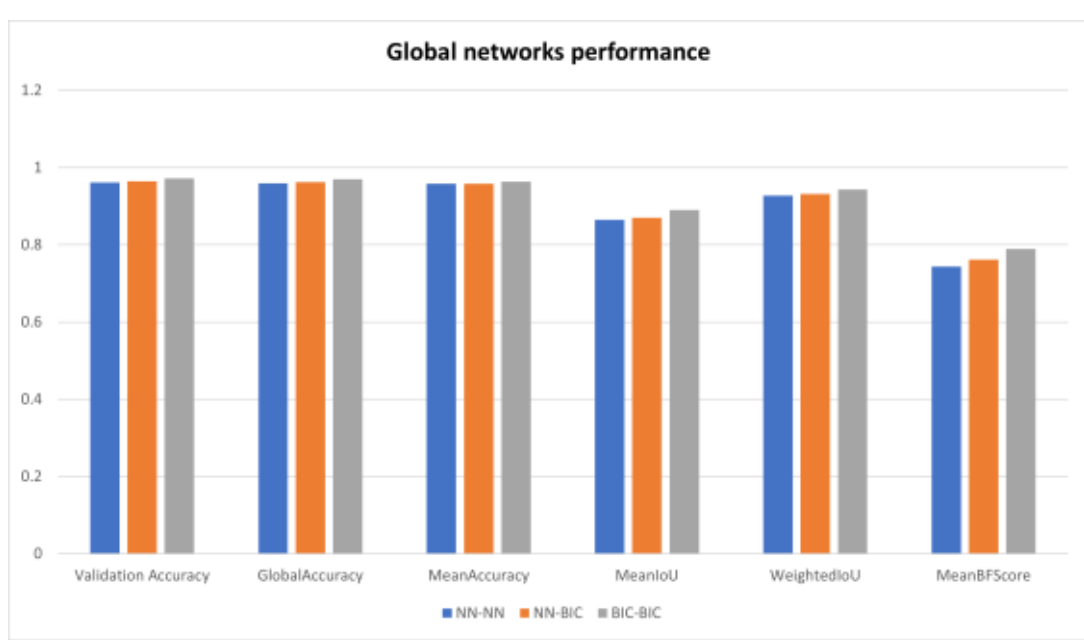

**Figure 10:** Global performance of networks trained with datasets consisting of 384 × 384 images and masks: NN-NN, BIC-NN, and BIC-BIC.

### 3.6. Discussions

In this section, the percentage increase is used to provide a clear and concise representation of the progress made by extra pixel category interpolation-based networks (BIC-NN or BIC-BIC) compared to the initial values or scores achieved by the non-extra pixel category interpolation-based network (NN-NN). Equation 1's output means that the extra pixel category interpolation-based network (A) is the increase of the non-extra pixel category interpolation-based network (B). Also, it means that the higher the percentage increase towards 100%, the better the score or the closer to the maximum score of 1.

$$\%\ increase = \frac{(A - B)}{B} \times 100 \qquad (1)$$

Table 1 and Table 2 show percentage increase rates in terms of Accuracy. Table 3 and Table 4 show percentage increase rates in terms of IoU. Table 5 and Table 6 show percentage increase rates in terms of meanBF. Table 7 shows percentage increase rates in terms of global accuracy. As can be seen, in Table 1, Table 2, Table 3, Table 4, Table 5, and Table 6, the BIC-BIC network achieved always a better average percentage increase than both NN-NN and BIC-NN networks.

**Table 1:** Extra pixel category interpolation-based networks percentage increase of NN-NN - case: **Accuracy** | size: **256 × 256**

| | **Region 1** | **Region 2** | **Region 3** |
|---|---|---|---|
| BIC-NN (2.5%) | -4.1334 % | 0.6599 % | **11.2193** % |
| BIC-BIC (7.0%) | **6.6446** % | **4.5382** % | 9.9760% |

**Table 2:** Extra pixel category interpolation-based networks percentage increase of NN-NN - case: **Accuracy** | size: **384 × 384**

| | Region 1 | Region 2 | Region 3 |
|---|---|---|---|
| BIC-NN(0.03%) | **1.0640** % | -1.6684 % | 0.4930 % |
| BIC-BIC (0.49%) | 0.4823 % | -0.3261 % | **1.3415** % |

**Table 3:** Extra pixel category interpolation-based networks percentage increase of NN-NN - case: **IoU** | size: **256 × 256**

| | Region 1 | Region 2 | Region 3 |
|---|---|---|---|
| BIC-NN (25.5%) | 51.0539 % | 16.0581 % | 9.63173 % |
| BIC-BIC (30.5%) | **63.4582** % | **18.2819** % | **9.9827** % |

**Table 4:** Extra pixel category interpolation-based networks percentage increase of NN-NN - case: **IoU** | size: **384 × 384**

| | Region 1 | Region 2 | Region 3 |
|---|---|---|---|
| BIC-NN (0.7%) | 2.5206 % | -0.5792 % | 0.4185 % |
| BIC-BIC (3.3%) | **8.3745** % | **0.3680** % | **1.2461** % |

**Table 5:** Extra pixel category interpolation-based networks percentage increase of NN-NN - case: **meanBF** | size: **256 × 256**

| | Region 1 | Region 2 | Region 3 |
|---|---|---|---|
| BIC-NN (55.7%) | 82.264 % | 48.4785 % | 36.515 % |
| BIC-BIC (69.5%) | **101.5649** % | **67.2667** % | **39.8251** % |

**Table 6:** Extra pixel category interpolation-based networks percentage increase of NN-NN - case: **meanBF** | size: **384 × 384**

| | Region 1 | Region 2 | Region 3 |
|---|---|---|---|
| BIC-NN (2.4%) | 4.5472 % | -2.5469 % | 5.3134 % |
| BIC-BIC (6.3%) | **12.5871** % | -2.1037 % | **8.5229** % |

**Table 7:** Extra pixel category interpolation-based networks percentage increase of NN-NN - case: **Global accuracy**

| | 256 × 256 case | 384 × 384 case |
|---|---|---|
| BIC-NN | 8.3127 % | 0.2887 % |
| BIC-BIC | **8.9578** % | **1.0496** % |

It is worth noting that the global accuracy was of interest here, but referring to Figure 10, it was clear that the BIC-BIC network outperformed the NN-NN and BIC-NN networks in terms of other cases, such as Validation Accuracy, MeanAccuracy, MeanIoU, WeightedIoU, and MeanBFScore for both 256 x 256 and 384 x 384 cases, respectively, thus suppressing the need for adding additional relevant percentages increase tables, in this part.

## 4. CONCLUSION

The use of extra pixel interpolation - with the interpolation mask processing strategy proposed in [6] - enhanced the accuracy of medical image segmentation in deep learning. New experimental results demonstrated that the mask processing-based extra pixel interpolation approach outperformed the conventional NN-based approach thus leading to improvements in segmentation accuracy. Specifically, the interpolation mask processing method enabled the BIC-BIC network to achieve superior segmentation accuracy, surpassing the performance of other combinations presented. Furthermore, experimental results also showed a higher percentage increase with dataset cases of image size 256 x 256 than 384 x 384. Further research may focus on improving data augmentation quality through the use of extra pixel interpolation and mask processing.

## DECLARATIONS

**Ethical Approval**
Consent to publish is granted.

**Competing interests**
There is no competing interest related to this work.

**Authors' contributions**
The author carried out 100% of the work presented.

**Funding**
Not applicable

**Availability of data and materials**
Data are not public but can only be made available upon request.